# Advanced protection against environmental degradation of silver mirror stacks for space application

*Rajendra Kurapati,[a] Vincent Maurice,[a] Antoine Seyeux, [a] Lorena H. Klein,[a] Dimitri Mercier,[a] Grégory Chauveau,[b] Catherine Grèzes-Besset,[b] Loïc Berthod,[c] Philippe Marcus[a]*

[a]PSL Research University, CNRS-Chimie ParisTech, Institut de Recherche de Chimie Paris, Physical Chemistry of Surfaces Group, 11 rue Pierre et Marie Curie, 75005 Paris, France

[b]CILAS MARSEILLE, 600 avenue de la Roche Fourcade, Pole ALPHA Sud - Saint Mitre, 13400 Aubagne, France

[c]CNES, Sous-direction Assurance Qualité, Service Technologies, Matériaux et Procédés, 18 Avenue Édouard Belin, 31401 Toulouse, France

## Abstract

Protection of silver mirror stacks from environmental degradation before launching is crucial for space applications. Hereby, we report a comparative study of the advanced protection of silver mirror stacks for space telescopes provided by $SiO_2$ and $Al_2O_3$ coatings in conditions of accelerated aging by sulfidation. The model silver stack samples were deposited by cathodic magnetron sputtering on a reference silica substrate for optical applications and a surface-pretreated SiC substrate. Accelerated aging was performed in dry and more severe wet conditions. Optical micrographic observations, surface and interface analysis by Time-of Flight Secondary Ion Mass Spectrometry (ToF-SIMS) and reflectivity measurements were combined to comparatively study the effects of degradation. The results show a lower kinetics of degradation by accelerated aging of the stacks protected by the alumina coating in comparable test conditions.

# Introduction

Environmental degradation in earth atmosphere is of major concern for the space industry since it is a limiting factor for the long-term durability of silver mirrors for demanding applications such as space telescopes. Local degradation by tarnishing of the reflective silver layer is observed during indoor qualification and storage phases after satellite assembly and before launching. It occurs despite the presence of protection and confinement coatings that have been developed to improve the durability of the mirrors [1-4]. The protective overcoats can consist of $SiO_2$ [5,6] $Al_2O_3$ [7,8] $Si_3N_4$ [9,10] or $SiN_x$ [11-13] dielectric layers deposited on the reflecting layer made of silver, this metal having the highest reflectivity in the visible and a low emissivity in the infrared [14,15].

Degradation studies performed on space silver mirrors mostly addressed the efficiency of the protection layer and the effect of environmental aging on the optical properties [8,13,16-19]. The tarnishing mechanism has also been studied [20-24]. In indoor atmospheres, the predominant tarnish product of silver metal is silver sulfide ($Ag_2S$) resulting from the corrosive interaction with organosulfur compounds or $H_2S$ pollutants [25-31]. In laboratory, $H_2S$ is mostly used to study silver sulfidation [16,32-36].

Recently, model stack samples consisting of thin silver layers protected by $SiO_2$ overcoats deposited by cathodic magnetron sputtering were studied on SiC [23,24], a substrate already in use in light weight all-SiC telescopes assigned to scientific or earth observation missions. Exposure to $H_2S$ gas was applied for accelerated aging and degradation was studied by surface and interface analysis. Local tarnishing was observed to result from the formation of $Ag_2S$ columns emerging above the stack surface. The growth of these $Ag_2S$ columns was promoted at the high aspect ratio surface defects (surface pores) of the SiC substrate due to imperfect protection by the $SiO_2$ overcoat. In the failure sites, permeation channels was proposed to connect the silver layer to the environment through the deposited protection layer, which would enable local $H_2S$ entry and $Ag_2S$ growth with eruption at the stack surface [23]. Suppressing the surfaces pores resulting from the bulk SiC material production process by surface pretreatment was found to eliminate the high aspect ratio surface sites imperfectly protected by the $SiO_2$ overcoat and thereby to markedly delay tarnishing initiation and longer preserve the optical performance [24].

Here we report a comparative study of the advanced protection provided by $SiO_2$ and $Al_2O_3$ overcoats in conditions of dry and more severe wet accelerated aging by sulfidation. The model silver stack samples were deposited by cathodic magnetron sputtering on a reference silica substrate for optical applications and a surface-pretreated SiC substrate. Accelerated aging was performed in dry and more severe wet conditions. Optical micrographic observations, surface and interface analysis by Time-of Flight Secondary Ion Mass Spectrometry (ToF-SIMS) and reflectivity measurements were combined to comparatively study the effects of degradation.

## Experimental

Substrate samples ($10\times10\times1.5$ mm$^3$) made of silica and Boostec® silicon carbide were bought from Mersen Boostec. The $SiO_2$ substrate samples have an optical surface finish with a typical roughness of less than 1 nm. The SiC CVD samples had one face pretreated by chemical vapor deposition at Mersen Boostec so as to deposit a ~200 µm thick high purity SiC layer in order to suppress the surface pores left by the material processing. After pretreatment, these SiC CVD substrate samples had an optical finish with a nominal roughness of 1.2 nm of the pretreated surface.

The PACA2M deposition platform at CILAS, in its premises of Aubagne [37-39] was used for preparation of the layered stacks by cathodic magnetron sputtering as in our previous studies [23,24]. Base pressure is $10^{-7}$ mbar and working pressure during deposition in the $10^{-3}$–$10^{-2}$ mbar range. The platform is equipped with an *in situ* broadband optical monitoring of the optical performance and thickness during deposition. Prior to deposition, the target cathode surfaces were pretreated by Ar sputtering in order to minimize their surface contamination. An adhesion interlayer consisting of metallic nickel and chromium and about 10 nm thick was first deposited on the substrates, then the silver mirror layer, 100 to 200 nm thick, then another adhesion interlayer (thickness and material of this interlayer cannot be disclosed), and finally a dielectric protection layer consisting of silicon dioxide or aluminum oxide and between 140 and 200 nm thick depending on the targeted optical performance. After coating deposition, the stack samples, taken out to clean room atmosphere, were enclosed in membrane boxes under ambient pressure before shipping to CNRS-Chimie ParisTech for accelerated aging and surface and interface analysis.

Dry accelerated aging was performed in $H_2S$ gas as detailed before [23,24]. The as-received stacks were mounted into a quartz tube devoted to thermal treatments under reactive gases. Sample edges were not protected. At first, the tube was pumped to secondary vacuum in the $10^{-5}$ mbar range, and then filled with $H_2S$ gas (5N purity, Air Liquide) at a pressure of 1000 mbar. Accelerated sulfidation was performed at 75°C using a tubular oven and during 96 h. Thermal treatment was stopped by quenching the quartz tube with iced water and simultaneously pumping back the system to secondary vacuum. In $H_2S$ gas, the formation of silver sulfide ($Ag_2S$) results from the dissociative adsorption of $H_2S$ on the silver metal surface unprotected by the dielectric coating, i.e. at the end of the permeation channels connecting the silver layer to the environment and bypassing the protection layer [23].

Wet accelerated aging was performed in a non-deaerated aqueous solution of 1 mM $Na_2S$ and 0.1 M NaOH of pH 13, prepared from reagent grade chemicals and ultra-pure water. At this pH, all (99.9%) sulfide species are $HS^-$ ions, the pKa values of the $H_2S/HS^-$ and $HS^-/S_2^-$ couples being 7.0 and 17.1, respectively [40]. An aqueous solution of pH 7 was also prepared by adding 0.1 M $H_2SO_4$ to the pH 13 solution, alumina being amphoretic and only stable in the 4 < pH < 8.3 range. At this pH, the solution is less aggressive since containing less (50%) HS⁻ ions, the other sulfide species being $H_2S$ (50%). Immersions were performed at free potential, silver sulfide being formed spontaneously in these solutions without applying any anodic polarization [41-44]. Immersion tests were performed for 2, 4, 6 and 8 hours. The immersed area was 0.28 cm$^2$ delimited by a Viton O ring, thus preserving the sample edges from exposure to the aggressive solution. In these wet conditions, the formation of silver sulfide ($Ag_2S$) results from the dissociative adsorption on the silver metal surface of the HS⁻ ions also permeating the protection layer via the connected porosity of the dielectric coating.

Optical microscope images were taken before and after sulphidation for all tested samples. Surface and interface analyses were performed by ToF-SIMS on selected samples. A IonTof 5 spectrometer operating at a base pressure of approximately $10^{-9}$ mbar was used. Chemical imaging was performed in static SIMS conditions. A pulsed 25 keV Bi⁺ primary ion source was employed, delivering 0.15 pA of current over a $100 \times 100$ μm$^2$ area analyzed with a resolution of about 150 nm. For depth profiling analysis, static SIMS analytical conditions were also used with a pulsed 25 keV Bi⁺ primary ion source delivering a target current of 1.2 pA over a $100 \times 100$ μm$^2$ area. It was interlaced with sputtering performed with another gun of 2 keV Cs⁺ ions and delivering 100 nA of target current over a $300 \times 300$ μm$^2$ area. Analysis in static SIMS conditions was centered in the sputtered crater to avoid edge effects. Data acquisition and post-processing analyses were performed with the Ion-Spec software.

## Results and Discussion

### *Resistance to dry accelerated aging on silica substrate*

Figure 1 compares optical images before and after 96 h of dry accelerated aging for the two stacks deposited on the silica substrate. On both stacks, degradation occurs predominantly by forming a delamination strip propagating from the sample edges, as expected since the samples chamfers were not polished and the edges were not masked during deposition. The delamination strip is narrower on the sample with the alumina overcoat, showing less advanced corrosion. Further away from the edges and in the sample center, corrosion spots with a maximum size reaching 25 µm are observed on the sample with the silica overcoat (Figure 1, top row). On the sample with the alumina overcoat (Figure 1, bottom row), smaller corrosion spots (maximum size reaching 5 µm) are observed near the sample edge and none in the sample center, thus showing that the alumina overcoat provides higher corrosion protection both at the sample edges and in regions not affected by edge effects.

Figure 2 shows ToF-SIMS chemical maps measured locally (analyzed area of $100 \times 100$ µm$^2$) on both samples after the dry accelerated aging test. The images were recorded at the topmost surface of the stacks after a slight sputtering to remove the surface contamination from exposure to ambient air. They are therefore chemical maps of the extreme surface of the protection layers. On the sample with the silica overcoat (Figure 2(a)), the selected negative ions are representative for the protection layer ($O^-$, $SiO_2^-$), for formation of silver ($^{109}Ag^-$) sulphide ($^{34}S^-$) and for contamination ($Cl^-$). The total ions map sums up the selected ions and others ($^{18}O^-$, $Ag^-$, $S^-$) not shown. Several (six) defects (dark spots) with a size less than 20 µm are present at the overcoat surface, some of them being aggregated. All these defects except the smallest coincide with the local formation of silver sulfide spots, as confirmed by the total ions maps that superimposes the silver sulfide spots inside the overcoat surface defects. The Cl contamination is quite homogeneous with a slight accumulation in the largest overcoat surface defect. The presence of sulfide corrosion spots inside the overcoat defects at the extreme surface of the stack is consistent with the mechanisms previously proposed for localized degradation with pre-existing channels (connected porosity) in the protection layer enabling $H_2S$ penetration down to the silver layer followed by local growth of silver sulfide with columns emerging at the stack surface [23,24].

On the sample with the alumina overcoat (Figure 2(b)), one defect, about 5 µm in size, is also observed at the overcoat surface (O$^-$ and SiO$_2^-$ ions maps). However, it is not associated with silver sulfide spots. The local presence of sulfur is detected in the S$^-$ map (not shown) but not in the $^{34}$S$^-$ isotopic map of lower intensity. Cl contamination is present in the defect at the same level as in the protection layer. Thus, although a surface defect is present in the protection layer, it may be a preferential pathway for local penetration of $H_2S$ but not for subsequent local growth of silver sulfide with emergence at the stack surface. Although the chemical mapping is local and may not be statistically representative of the whole sample, the data are in coherence with the optical imaging analysis showing that the alumina overcoat seems to provide better protection against environmental aging with less and smaller pre-existing defects (lower connected porosity) for preferential penetration of $H_2S$ triggering local silver sulfidation on these samples.

Figure 3 shows the ToF-SIMS elemental depth profiles measured on both samples after the dry accelerated aging test. The intensity of the selected negative ions is plotted in logarithmic scale versus sputtering time. Starting from the outer surface, the intensity profiles define the main three regions of the deposited stacks corresponding successively to the protection layer ($^{18}$O$^-$, SiO$_2^-$ and SiO$^-$ ions in Figure 3(a) and $^{18}$O$^-$, AlO$_2^-$ and AlO$^-$ ions in Figure 3(b)), the silver layer (Ag$^-$ ions) and the SiO$_2$ substrate ($^{18}$O$^-$, SiO$_2^-$ and SiO$^-$ ions). The observed variations of sputtering time between the SiO$_2$/Ag/SiO$_x$ and SiO$_2$/Ag/AlO$_x$ stacks comes from the fact that they were produced in different batches with variations of the deposited thickness of the silver layer (from 100 to 200 nm thick) and protective layer (from 140 to 200 nm). The adhesion interlayer regions $I_a$ and $I_b$ are marked by a peak and a steep decrease in the Ag$^-$ ions profile, respectively. The peak of the Ag$^-$ ions profile in the $I_a$ interfacial region is consistent with the formation of a discontinuous adhesion interlayer forming islands as discussed previously [23,24]. In the $I_b$ interfacial region, the expected $Ag^-$ ions intensity drop down coincides with the expected steep increases in the $^{18}O^-$, $SiO_2^-$ and $SiO^-$ ions profiles also characteristic of the substrate.

On the sample with the silica overcoat (Figure 3(a)) and like previously observed [23,24], the $^{18}O^-$, $SiO_2^-$ and $SiO^-$ ions exhibit plateaus of increasing intensity in the protection layer region with in-depth progress, suggesting a density of the $SiO_2$ layer decreasing with on-going deposition from the $I_a$ interface. Silver contamination of the target cathode during previous deposition of the reflective layer could explain the detection of the $Ag^-$ ions in the overcoat region. This Ag presence could also result from photo-enhanced migration during deposition promoted by oxygen adsorption on silver, as previously proposed for a $TiO_2$ protection layer [45]. The profile of the $Cl^-$ ions, characteristic of atmospheric contamination by chlorine-containing pollutants, is consistent with the presence of channels in the protection layer exposing the underlying silver layer and substrate to the environment. This connected porosity is also supported by the $^{34}S^-$ ions profile, characteristic of the penetration of $H_2S$ during the accelerated aging test since the intensity was at noise level before the test. The formation of silver sulfide columns growing throughout the protection layer and emerging at the stack surface is confirmed by the $AgS^-$ ions profile of low but non-zero intensity in the overcoat region. Hence the ToF-SIMS profiles provide indirect evidence of the presence of permeation channels penetrating the coating and at the origin of the local growth of the silver sulfide corrosion spots.

On the sample with the alumina overcoat (Figure 3(b)), the $AlO_2^-$ and $AlO^-$ ions also exhibit plateaus of increasing intensity in the protection layer region with in-depth progress, also suggesting a density of the $Al_2O_3$ layer decreasing with on-going deposition from the $I_a$ interface. Silver contamination appears more homogeneous in depth than in the silica overcoat. Cl contamination is at similar level in the protection layer but also observed in the silver layer, suggesting that some may have occurred during deposition (these samples with different overcoats were produced in two separate runs). The $^{34}S^-$ ions profile has similar intensity at the surface of the protection layer, as expected from the exposure to $H_2S$ gas, but then drops near noise level with in-depth progress inside the alumina overcoat, which is characteristic of the reduced penetration of $H_2S$ during the accelerated aging test compared to the silica-coated sample. The $AgS^-$ ions profile of zero intensity in the overcoat region is consistent with the absence of formation of silver sulfide columns growing throughout the protection layer. Some intensity is only detected at the surface of the silver layer suggesting that silver sulfide growth has initiated but not propagated through the protective overcoat.

Thus in-depth profile analysis of the stacks after dry accelerated aging confirms the better protection provided by the alumina overcoat for the mirror stack deposited on the silica substrate. Penetration of $H_2S$ via the connected porosity of the protection layer would occur with both types of overcoats but would more slowly reach the silver layer with the alumina overcoat, suggesting a more compact coating with lower connected porosity delaying the initiation of localized degradation by sulfidation and the local growth of silver sulfide columns through the overcoat.

***Resistance to dry accelerated aging on SiC CVD substrate***

Figure 4 compares optical images before and after 96 h of dry accelerated aging for the two stacks deposited on the SiC CVD substrate. Preferential degradation is also observed at the sample edges with more advanced propagation of the delamination strip on the sample with the silica overcoat. In the samples center regions not affected by edge effects, one also observes larger (largest size of ~25 µm vs ~10 µm) corrosion spots on the sample with the silica overcoat but the spots appear more numerous on the sample with the alumina overcoat. These observations suggest a higher resistance to propagation with the alumina overcoat for the stacks deposited on the SiC CVD substrate. The higher probability of initiation of localized degradation could be either due a different amount of localized defects on the SiC CVD substrate or a lower resistance to the initiation of localized degradation.

Figure 5 shows ToF-SIMS chemical maps measured locally on both samples after the dry accelerated aging test. On the sample with the silica overcoat (Figure 5(a)), one defect (dark spot) with a size less than 10 µm is observed at the overcoat surface in the analyzed area. Its presence coincides with the local formation of a silver sulfide spot, as confirmed by the $^{109}Ag^-$, $^{34}S^-$ and total ions maps. The Cl contamination is quite homogeneous with no accumulation in the overcoat surface defect. On the sample with the alumina overcoat (Figure 5(b)), several defects with a size less than 5 µm are observed at the overcoat surface, three of them being aggregated to form a larger one. They are also associated with the local formation of a silver sulfide spots except for one of them, as confirmed by the $^{109}Ag^-$, $^{34}S^-$ and total ions maps. Cl contamination is also observed where sulfide spots are formed. Thus, and although the chemical mapping analysis may not be statistically representative of the whole sample behavior, it is confirmed that the sample with the alumina overcoat would be more prone to initiation of degradation by local sulfidation but better resist propagation by growth of the silver sulfide corrosion spots for the stacks deposited on the SiC CVD substrate.

The ToF-SIMS elemental depth profiles measured on both samples after the dry accelerated aging test are shown on Figure 6. On the sample with the silica overcoat (Figure 6(a)), the profile of the $Cl^-$ ions, characteristic of atmospheric contamination by chlorine-containing pollutants, is consistent with the presence of channels in the protection layer exposing the underlying silver layer and substrate to the environment, as for the stack deposited on the $SiO_2$ substrate (Figure 3(a)). The penetration of $H_2S$ via this connected porosity during the accelerated aging test is also supported by the intensity of the $^{34}S^-$ ions profile, higher than the noise level measured before the test. The few counts of $AgS^-$ ions profile in the overcoat region is consistent with silver sulfide columns growing throughout the protection layer. Again these ToF-SIMS profiles provide indirect evidence of the presence of permeation channels bypassing the protection provided by the dielectric overcoat.

On the sample with the alumina overcoat (Figure 6(b)), Cl contamination is observed in the protection layer and in the silver layer like for the stack deposited on the $SiO_2$ substrate (Figure 3(a)), confirming that it may have occurred during deposition. The $^{34}S^-$ ions profile has similar intensity at the surface of the protection layer but remains at a higher level with in-depth progress inside the alumina overcoat than for the stack deposited on the $SiO_2$ substrate (Figure 3 (b)), which is indicative of the deeper penetration of $H_2S$ during the accelerated aging test down to the silver layer. The $AgS^-$ ions profile is barely above noise level in the outer part of the overcoat region but has higher intensity in the inner part, in the $I_a$ interface region and in the silver layer region than for the stack deposited on the $SiO_2$ substrate. This supports the faster initiation of the growth of silver sulfide owing to a higher, i.e. more connected, porosity of the alumina coating deposited on the SiC CVD substrate.

Compared with the sample with the silica overcoat (Figure 6(a)), the $^{34}S^-$ ions profile is also consistent with a connected porosity penetrating throughout the protection layer and reaching the silver layer. The $AgS^-$ ions profile is not detected in the outer part of the overcoat region but has higher intensity in the inner part and in the silver layer region, also indicating the lower resistance to the initiation of local sulfidation but higher resistance to the growth of the silver sulfide columns inferred from the micrographic observations.

These differences are consistent with the observation of a lower resistance of the protected silver stacks to the initiation of local sulfidation when deposited on the SiC CVD substrate than on the $SiO_2$ substrate. This substrate effect might find its origin in the surface pretreatment of the SiC CVD substrate which was found necessary to markedly improved the resistance to localized sulfidation of the stacks deposited on this substrate [23,24]. Any variability in the surface pretreatment for healing the surface defects might impact the quality of the deposited stacks and their resistance to local sulfidation during accelerated aging.

### *Resistance to wet accelerated aging on SiC CVD substrates*

Figure 7 compares optical images before and after 2, 4, 6 and 8 h of wet accelerated aging for the two stacks deposited on the SiC CVD substrate. Two regions observed inside the exposed area are shown in the center and right columns for each specimen. On the samples with the silica overcoat (Figure 7(a)), all immersion tests were conducted with the more aggressive pH 13 solution. Corrosion spots are observed to increase in number and size with increasing immersion time showing delayed initiation of the smaller spots and growth of the larger spots. Their disk shape is indicative of radial growth around the initiation point. After 6 hours of treatment, the corrosion spots have a maximum size of about 70 µm. On the sample treated for 8 hours, less numerous and smaller spots are observed suggesting less advanced corrosion that we assign to some disparities between substrate samples owing to the surface preparation of the substrate. On this particular sample, the silica overcoat stack would contain less and/or smaller channels for the penetration of the sulfide species (HS⁻) and thus better protect from the initiation and growth of the corrosion spots.

On the samples with the alumina overcoat (Figure 7(b)), the immersion test conducted for 2 hours was also performed with the more aggressive pH 13 solution. The sample is fully corroded because of the instability of alumina that has dissolved at this alkaline pH, confirming that alumina cannot offer durable protection in these conditions because it is amphoteric. The tests conducted for 4, 6 and 8 hours were conducted with the less aggressive pH 7 solution. In these conditions, alumina is stable and protects the silver layer against generalized corrosion. As with the silica overcoat at pH 13, only localized corrosion is observed with formation of corrosion spots. The corrosion spots are smaller (maximum size of about 15 µm) but not significantly less numerous than those observed with the silica overcoat at pH 13, a trend also observed after dry accelerated aging in the $H_2S$ gas, suggesting that their growth rather than their initiation would be predominantly delayed. This difference may be related to the lower concentration in $HS^-$ species of the pH 7 solution. With increasing immersion time, one does not observe a marked increase of the corrosion spots in size and in number, indicating a better resistance of the alumina overcoat to growth than to initiation of localized corrosion, as observed after dry accelerated aging.

Figure 8 shows the specular reflectance spectra of the less resistant stacks deposited on the SiC CVD substrate prior to and after wet accelerated aging in the more aggressive aqueous solutions. On the samples with the silica overcoat (Figure 8(a)), all curves show typical reflectivity drops around 375 nm caused by the protection layers. Comparing the samples before and after aging shows the slight degradation of the optical performance. The loss of reflectivity caused by the development of the corrosion spots does not exceed an average value of 6 % on the most corroded sample (after 6 h aging) and of 2% on the other corroded samples, showing that reflectivity is well-preserved despite the quite severe conditions of the aging test.

On the samples with the alumina overcoat (Figure 8(b)), a slight different shape of reflectivity curve is observed due to a different protecting layer. Comparing the samples before and after aging shows no significant degradation of the optical performance. All curves superimpose in the 350 nm to 2 µm range, with some discrepancies due to the spectrophotometer accuracy. These data show that such SiC CVD samples with the stack protected by the alumina coating exhibit an excellent resistance to accelerated aging.

## Conclusion

Advanced protection against environmental degradation provided to silver mirrors for space telescopes by $SiO_2$ and $Al_2O_3$ overcoats was studied on stack model samples deposited by cathodic magnetron sputtering on two types of substrates, a reference silica substrate commonly used in optical applications and a CVD surface-pretreated SiC substrate. Accelerated aging by sulfidation was performed in dry ($H_2S$ gas, 1000 mbar, 75°C) and more severe wet ($HS^-$ ions-containing aqueous solution of pH 7 or 13) conditions.

For the stacks deposited on the silica substrate and submitted to dry accelerated aging, the alumina overcoat was found to provide higher protection against delamination at the sample edges and against localized sulfidation in regions not affected by edge effects. The optical micrography and ToF-SIMS data are consistent with the mechanism previously proposed for localized degradation with pre-existing channels (connected porosity) in the protection layer enabling $H_2S$ penetration down to the silver layer followed by local growth of silver sulfide with columns emerging at the stack surface. The lower connected porosity in the alumina overcoat would increase the barrier property, thereby delaying the initiation of localized degradation by sulfidation and the local growth of silver sulfide columns through the overcoat.

For the stacks deposited on the surface-pretreated SiC substrate, higher resistance to propagation of local silver sulfide growth was confirmed with the alumina protection. However, variability in the surface pretreatment might be at the origin of the lower resistance observed in the initiation of localized degradation, with a more connected porosity developed above the substrate surface sites imperfectly healed by the pretreatment.

In the more severe wet accelerated aging conditions, the alumina overcoat provides protection against sulfidation only at pH 7 where it is stable but not at pH 13 where it dissolves. The same trend of lower resistance to the initiation of local sulfidation but higher resistance to the growth of the silver sulfide columns was observed for the stacks more prone to local degradation deposited on the surface-pretreated SiC substrate. Specular reflectivity measurements confirmed the advanced protection provided by the overcoats. With the alumina overcoat, no significant loss of the optical performance was observed in the whole spectral range after the most severe aging test at pH 7. With the silica overcoat, the loss of the optical performance was limited to a few percent of reflectivity after aging at pH 13.

## Acknowledgment

Region Ile-de-France is acknowledged for partial funding of the ToF-SIMS equipment.

# References

[1] D.-Y. Song, R. W. Sprague, H. A. Macleod, M. R. Jacobson, Progress in the development of a durable silver-based high-reflectance coating for astronomical telescopes, Appl. Opt. 24 (1985) 1164–1170.

[2] M. R. Jacobson, R. C. Kneale, F. C. Gillett, K. Raybould, J. F. Filhaber, C. K. Carniglia, R. Laird, D. Kitchens, R. P. Shimshock, D. C. Booth, Development of silver coating options for the Gemini 8-m telescopes project, In: Proc. SPIE 3352, 477-502 (1998).

[3] M. Boccas, T. Vucina, C. Araya, E. Vera, C. Ahhee, Protected silver coatings for the 8-m Gemini telescope mirrors, Thin Solid Films 502 (2006) 275–280.

[4] D. A. Sheikh, Improved silver mirror coating for ground and space-based astronomy, In Proc. SPIE 9912, 991239 (2016).

[5] D.W. Rice, R. J. Cappell, W. Kinsolving, J. J. Laskowski, Indoor Corrosion of Metals, J. Electrochem. Soc. 127 (1980) 891-901.

[6] M. G. Dowsett, A. Adriaens, M. Soares, H. Wouters, V. V. N. Palitsin, R. Gibbons, R. J. H. Morris, The use of ultra-low-energy dynamic SIMS in the study of the tarnishing of silver, Nuclear Instruments and Methods in Physics Research Section B : Beam Interactions with Materials and Atoms 239 (2005) 51-64.

[7] G. Hass, Reflectance and preparation of front-surface mirrors for use at various angles of incidence from the ultraviolet to the far infrared, J. Opt. Soc. Am. 72 (1982) 27-39.

[8] G. I. N. Waterhouse, G. A. Bowmaker, J. B. Metson, Oxidation of a polycrystalline silver foil by reaction with ozone, Appl. Surf. Sci. 183 (2001) 191-204.

[9] J. M. Bennett, J. L. Stanford, E. J. Ashley, Optical constants of silver sulfide tarnish films, J. Opt. Soc. Am. 60 (1970) 224-231.

[10] A. C. Phillips, J. Miller, W. Brown, D. Hilyard, B. Dupraw, V. Wallace, D. Cowley, Progress toward high-performance reflective and anti-reflection coatings for astronomical optics, In Proc. SPIE 7018, 70185A (2008).

[11] J. D. Wolfe, R. E. Laird, C. K. Carniglia, Durable silver-based antireflection coatings and enhanced mirrors, In: Optical Interference Coatings, Vol. 17, OSA Technical Digest Series (Optical Society of America), 115–117 (1995).

[12] J. D. Wolfe, D. M. Sanders, S. Bryan, N. L. Thomas, Deposition of durable wide-band silver mirror coatings using long-throw, low-pressure, DC-pulsed magnetron sputtering,

In Proc. of SPIE: Astronomical Telescopes and Instrumentation: Specialized optical developments in astronomy, The International Society for Optics and Photonics, 343-351 (2003).

[13] D. A. Sheikh, S. J. Connell, R. S. Dummer, Durable silver coating for Kepler Space Telescope primary mirror, In Proc. SPIE 7010, 70104E (2008).

[14] J. M. Bennett, E. J. Ashley, Infrared Reflectance and Emittance of Silver and Gold Evaporated in Ultrahigh Vacuum, Appl. Opt. 4 (1965) 221-224.

[15] N. Thomas, J. Wolfe, UV-shifted durable silver coating for astronomical mirrors, in Optical Design, Materials, Fabrication, and Maintenance, In Proc. SPIE 4003, Optical Design, Materials, Fabrication, and Maintenance, pp. 312-323 (2000).

[16] D. K. Burge, H. E. Bennett, E. J. Ashley, Effect of Atmospheric Exposure on the Infrared Reflectance of Silvered Mirrors With and Without Protective Coatings, Appl. Opt. 12 (1973) 42-47.

[17] Fuqua P. D.; Barrie J. D.; Optical properties and corrosion resistance of durable silver coatings. Mat. Res. Soc. Symp. Proc. 555 (1998) 85–90.

[18] C. T. Chu, P. D. Chaffee, C. J. Panetta, P. D. Fuqua, J. D. Barrie, Mixed Flowing Gas Testing of Silver Mirror Coatings, In: OSA Technical Digest (CD), Optical Interference Coatings, TuEPDP5 (2007).

[19] A. Feller, K. Nagaraju, O. Pleier, J. Hirzberger, P. J. Jobst, M. Schuermann, Reflectivity, polarization properties and durability of metallic mirror coatings for the European Solar Telescope, In: Proc. SPIE 8450, 84503U, 2012.

[20] Pellicori S. F. Scattering defects in silver mirror coatings. Appl. Opt. 19 (1980) 3096-3098.

[21] C.-T. Chu, P. D. Fuqua et J. D. Barrie, Corrosion characterization of durable silver coatings by electrochemical impedance spectroscopy and accelerated environmental testing, Appl. Opt. 45 (2006) 1583-1593.

[22] K. A. Folgner, C.-T. Chu, Z. R. Lingley, H. I. Kim, J.-M. Yang, J. D. Barrie, Environmental durability of protected silver mirrors prepared by plasma beam sputtering. Appl. Opt. 56 (2017) C75-C86.

[23] E. Limam, V. Maurice, A. Seyeux, S. Zanna, L. H. Klein, G. Chauveau C. Grèzes-Besset, I. Savin De Larclause, P. Marcus, Local degradation mechanism by tarnishing of

protected silver mirror layers studied by combined surface analysis, J. Phys. Chem. B 122 (2018) 578-586.

[24] E. Limam, V. Maurice, A. Seyeux, S. Zanna, L. H. Klein, G. Chauveau C. Grèzes-Besset, I. Savin De Larclause, P. Marcus, Role of SiC substrate surface on local tarnishing of deposited silver mirror stacks, App. Surf. Sci. 436 (2018) 1147–1156.

[25] C. Leygraf, I. O. Wallinder, J. Tidblad, T. Graedel, *Atmospheric Corrosion, Second Edition*. John Wiley & Sons, Inc.: Hoboken, USA, 2016.

[26] H. A. Ankersmit, N. H. Tennent, S. F. Watts, Hydrogen sulfide and carbonyl sulfide in the museum environment- Part 1, Atmospheric Environment 39 (2005) 695-707.

[27] J. Horvath, L. Hackl, Check of the potential/pH equilibrium diagrams of different metal-sulphur-water ternary systems by intermittent galvanostatic polarization method, Corrosion Sci. 5 (1965) 525-538..

[28] D. Rice, R. Cappell, P. Phipps, P. Peterson, W. Ailor, Indoor atmospheric corrosion of copper, silver, nickel, cobalt and iron. In: Atmospheric Corrosion (W. H. Ailor, Ed.), Wiley: New York, pp. 651-666 (1982).

[29] Y. Fukuda, T. Fukushima, A. Sulaiman, I. Musalam, L. C. Yap, L. Chotimongkol, S. Judabong, A. Potjanart, O. Keowkangwal, K. Yoshihara, M. Tosa, Indoor Corrosion of Copper and Silver Exposed in Japan and ASEAN1 Countries, J. Electrochem. Soc. 138 (1991) 1238-1243.

[30] S. Bouquet, C. Bodin, et C. Fiaud, Relative Infuence of Sulfide and Chloride Compounds on Tarnishing of Silver in Atmospheric Corrosion, CR. Acad. Sci. II 316 (1993) 459-464.

[31] D. Liang, H. C. Allen, G. S. Frankel, Z. Y. Chen, R. G. Kelly, Y. Wu et B. E. Wyslouzil, Effects of Sodium Chloride Particles, Ozone, UV, and Relative Humidity on Atmospheric Corrosion of Silver, J. Electrochem. Soc. 157 (2010) C146-C156.

[32] D. Pope, H. R. Gibbens, R. L.Moss, The tarnishing of Ag at naturally-occurring H2S and SO2 levels, Corrosion Sci. 8 (1968) 883-887.

[33] S. Lilienfeld, C. E. White, A Study of the Reaction Between Hydrogen Sulfide and Silver. J. Am. Chem. Soc. 52 (1930) 885-892.

[34] S. Kasukabe, Growth mechanism and growth form of Beta-Ag2S whiskers, J. Cryst. Growth 65 (1983) 384-390.

[35] T. E. Graedel, J. P. Franey, G. J. Gualtieri, G. W. Kammlott, D. L. Malm, On the mechanism of silver and copper sulfidation by atmospheric H2S and OCS, Corrosion Sci. 25 (1985) 1163-1180.

[36] D. W. Rice, P. Peterson, E. B. Rigby, P. B. P. Phipps, R. J. Cappell, R. Tremoureux, Atmospheric Corrosion of Copper and Silver, J. Electrochem. Soc. 128 (1981) 275-284.

[37] G. Chauveau, D. Torricini, C. Grèzes-Besset, D. Stojcevski, M. Lequime, PACA2M: magnetron sputtering for 2-meter optics. In Proc. SPIE 8168, 81680P (2011).

[38] M. Lequime, C. Grezes-Besset, G. Chauveau, D. Stojcevski, Optimization of the Manufacturing Strategies of High Quality Coatings into a 2-Meter Optics Magnetron Sputtering Deposition Machine. In 2014 Technical Conference Proceedings, Optical Coatings (September 29, 2014) TechCon2014, Society of Vacuum Coaters (2014).

[39] N. Valette, G. Chauveau, C. Grèzes-Besset, V. Costes, I. Savin de Larclause, K. Gasc, F. Lemarquis, PACA2M Magnetron sputtering silver coating: a solution for very big mirror dimensions, In ICSO 2014 Proceedings, International Conference on Space Optics, Tenerife, Canary Islands, Spain, 7 - 10 October 2014.

[40] W. Giggenbach, Optical spectra of highly alkaline sulfide solutions and the second dissociation constant of hydrogen sulfide, Inorganic Chemistry 10 (1971) 1333-1338.

[41] V. I. Birss, G. A.Wright, The kinetics of the anodic formation and reduction of phase silver sulfide films on silver in aqueous sulfide solutions, Electrochimica Acta 26 (1981) 1809-1817.

[42] D. W. Hatchett, H. S. White, Electrochemistry of sulfur adlayers on the low index faces of silver, Journal of Physical Chemistry 100 (1996) 9854-9859.

[43] G. D. Aloisi, M. Cavallini, M. Innocenti, M. L. Foresti, R. Pezzatini, R. Guidelli, In situ STM and electrochemical investigation of sulfur oxidative underpotential deposition on Ag (111), Journal of Physical Chemistry B 101 (1997) 4774-4780.

[44] N. Li, V. Maurice, L. H. Klein, P. Marcus, Structure and Morphology Modifications of Silver Surface in the Early Stages of Sulfide Growth in Alkaline Solution, Journal of Physical Chemistry C 116 (2012) 7062-7072.

[45] K. Chiba, K. Nakatani, Photoenhance migration of silver atoms in transparent heat mirror coatings, Thin Solid Films 112 (1984) 359-367.

**Figure captions**

*Figure 1 Optical images of multilayer stacks $SiO_2/Ag/SiO_x$ (top row) and $SiO_2/Ag/AlO_x$ (bottom row) before (0 h) and after (96 h) dry accelerated aging in $H_2S$ gas at 1000 mbar and 75 °C. The scale bar represents 100 µm. Center and right column images were recorded in the center and at the edges of the samples, respectively.*

*Figure 2 (color online) ToF-SIMS chemical maps ($100 \times 100\,µm^2$) of selected negative secondary ions for multilayer stacks $SiO_2/Ag/SiO_x$ (a) and $SiO_2/Ag/AlO_x$ (b) after 96 h of dry accelerated aging in $H_2S$ gas at 1000 mbar and 75 °C. The images were recorded in the sample center region, away from the edges.*

*Figure 3 (color online) ToF-SIMS elemental depth profiles of selected negative secondary ions for multilayer stacks $SiO_2/Ag/SiO_x$ (a) and $SiO_2/Ag/AlO_x$ (b) after 96 h of dry accelerated aging in $H_2S$ gas at 1000 mbar and 75 °C. The profiles were recorded in the sample center region, away from the edges. The protection layer, silver (Ag) layer and substrate regions are marked, as well as the interfacial $I_a$ and $I_b$ regions.*

*Figure 4 Optical images of multilayer stacks SiC CVD/Ag/$SiO_x$ (top row) and SiC CVD/Ag/$AlO_x$ (bottom row) before (0 h) and after (96 h) dry accelerated aging in $H_2S$ gas at 1000 mbar and 75 °C. The scale bar represents 100 µm. Center and right column images were recorded in the center and at the edges of the samples, respectively.*

*Figure 5 (color online) ToF-SIMS chemical maps ($100 \times 100\,µm^2$) of selected negative secondary ions for multilayer stacks SiC CVD/Ag/$SiO_x$ (a) and SiC CVD/Ag/$AlO_x$ (b) after 96 h of dry accelerated aging in $H_2S$ gas at 1000 mbar and 75 °C. The images were recorded near the sample center region, away from the edges.*

*Figure 6 (color online) ToF-SIMS elemental depth profiles of selected negative secondary ions for multilayer stacks $SiO_2/Ag/SiO_x$ (a) and $SiO_2/Ag/AlO_x$ (b) after 96 h of dry accelerated aging in $H_2S$ gas at 1000 mbar and 75 °C. The profiles were recorded near the sample center region, away from the edges. The protection layer, silver (Ag) layer and substrate regions are marked, as well as the interfacial $I_a$ and $I_b$ regions.*

*Figure 7 Optical images of multilayer stacks $SiC$ $CVD/Ag/SiO_x$ (a) and $SiC$ $CVD/Ag/AlO_x$ (b) before (0 h) and after (2,4, 6 and 8 h) wet accelerated aging in 1 mM $Na_2S$ aqueous solution at pH 13 ((a) all rows and (b) first row) and pH 7 ((b) second to last row). The scale bar represents 100 µm. Center and right column images were recorded in two different sample locations.*

*Figure 8 (color online) Specular reflectance spectra for the stacks $SiC$ $CVD/Ag/SiOx$ (a) and $SiC$ $CVD/Ag/AlOx$ (b) before and after wet accelerated aging in 1 mM $Na_2S$ aqueous solution at pH 13 (a) and pH 7 (b).*

**Figure 1**

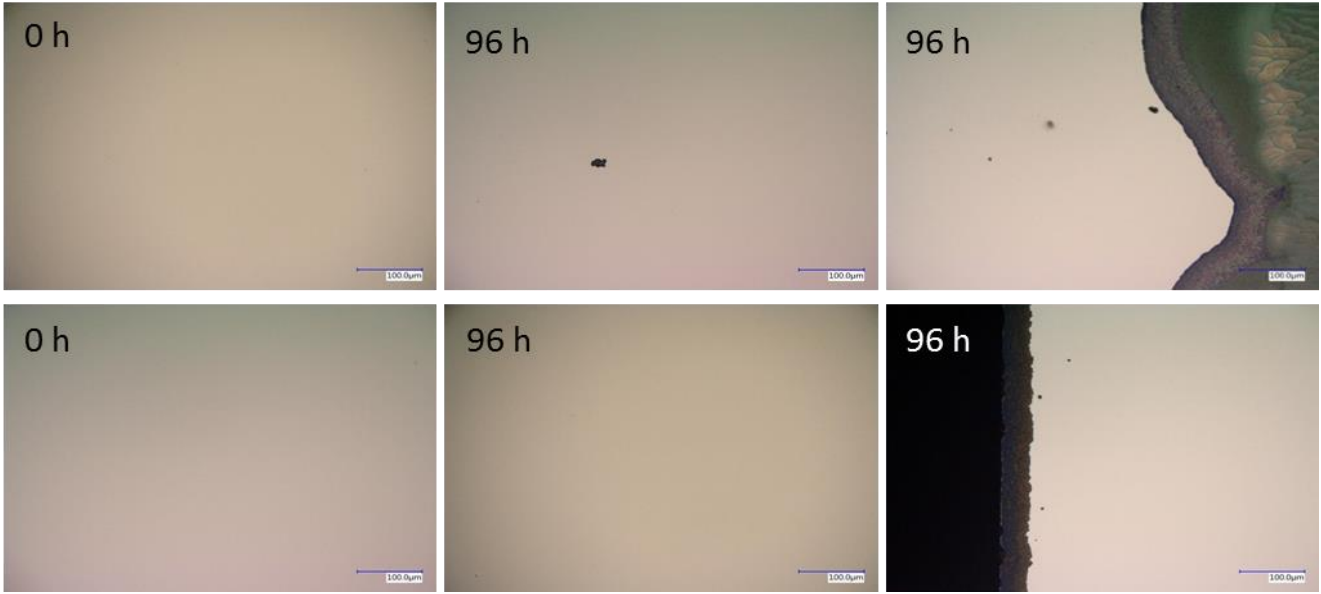

**Figure 2**

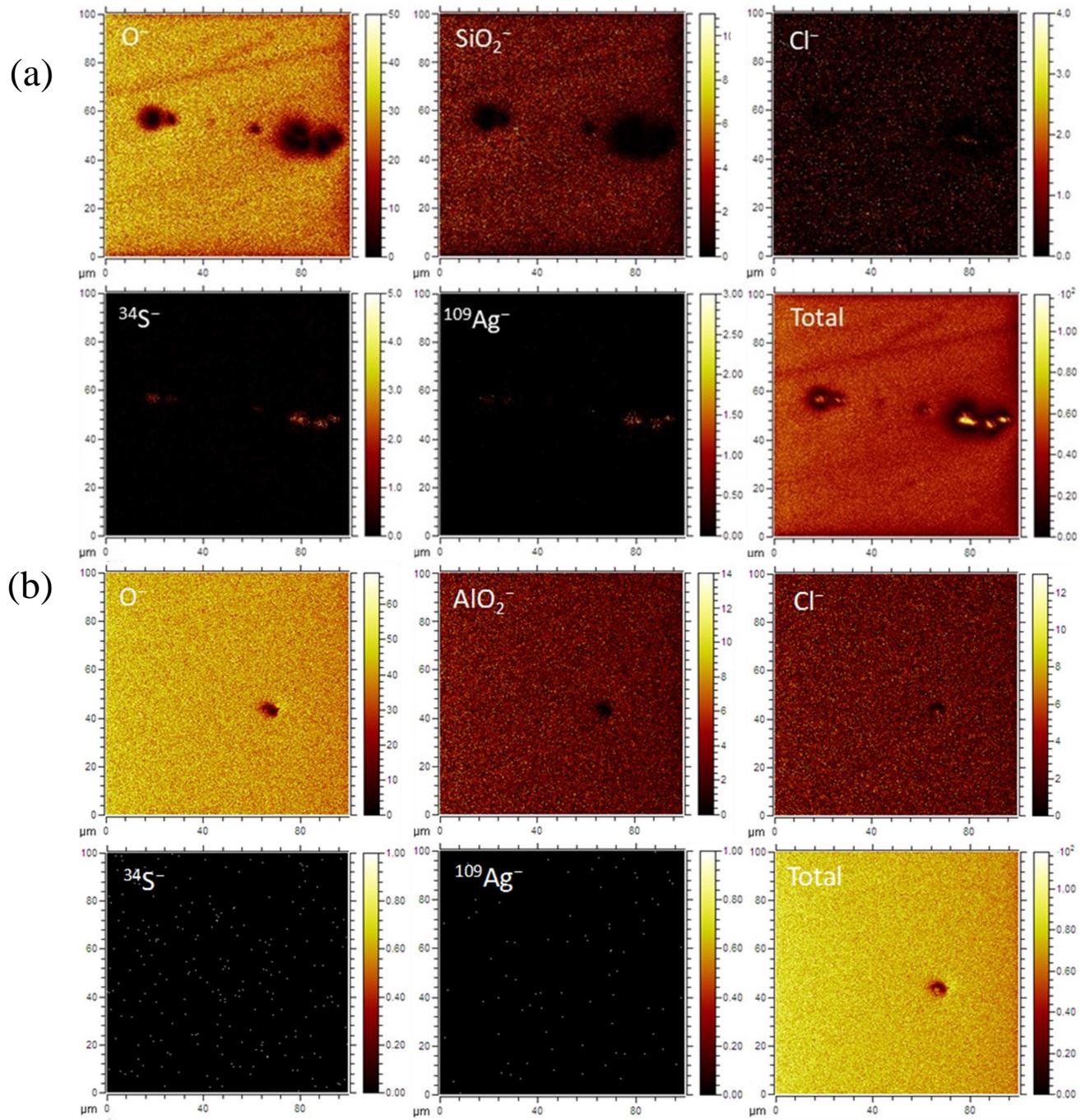

**Figure 3**

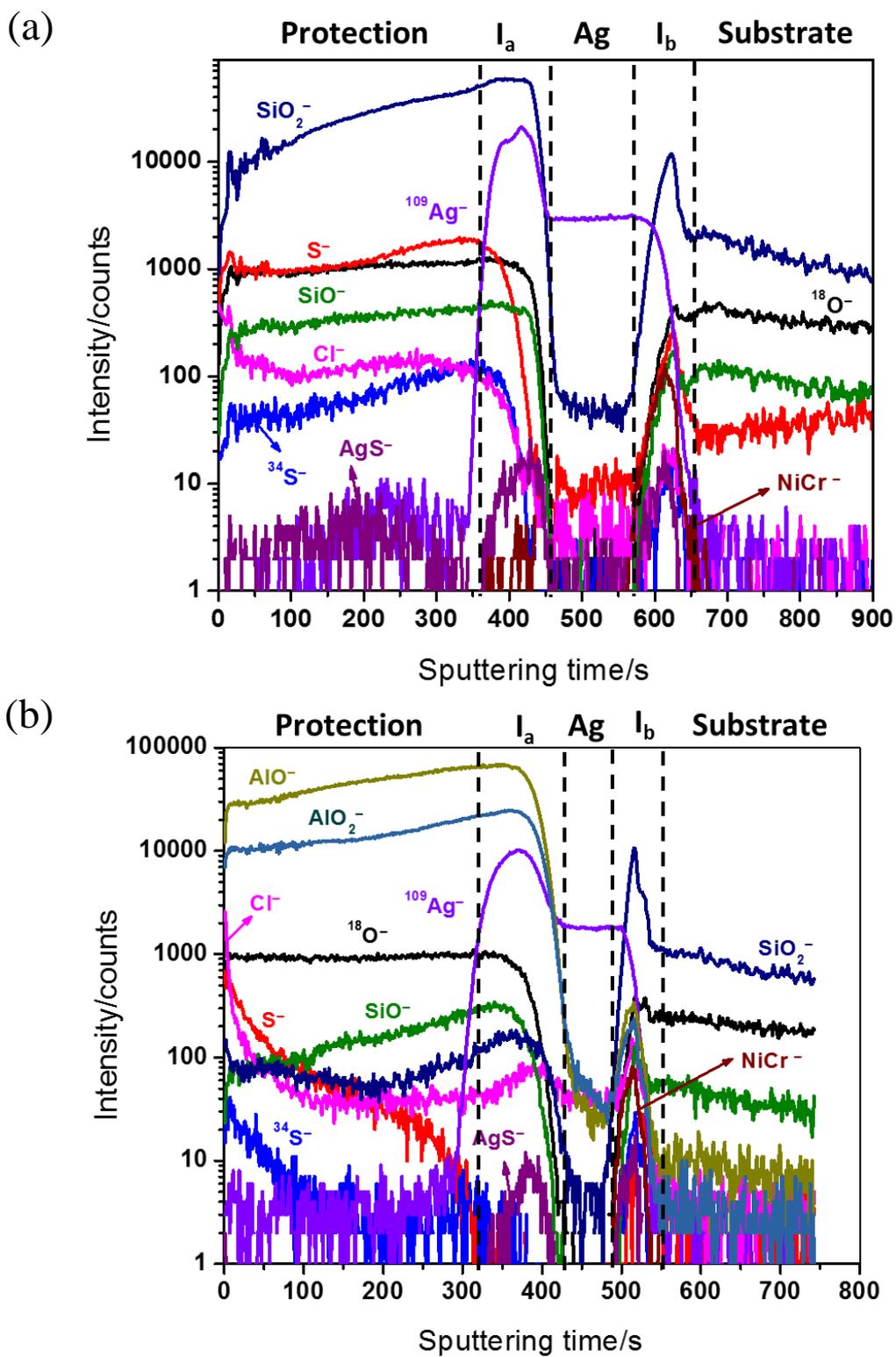

**Figure 4**

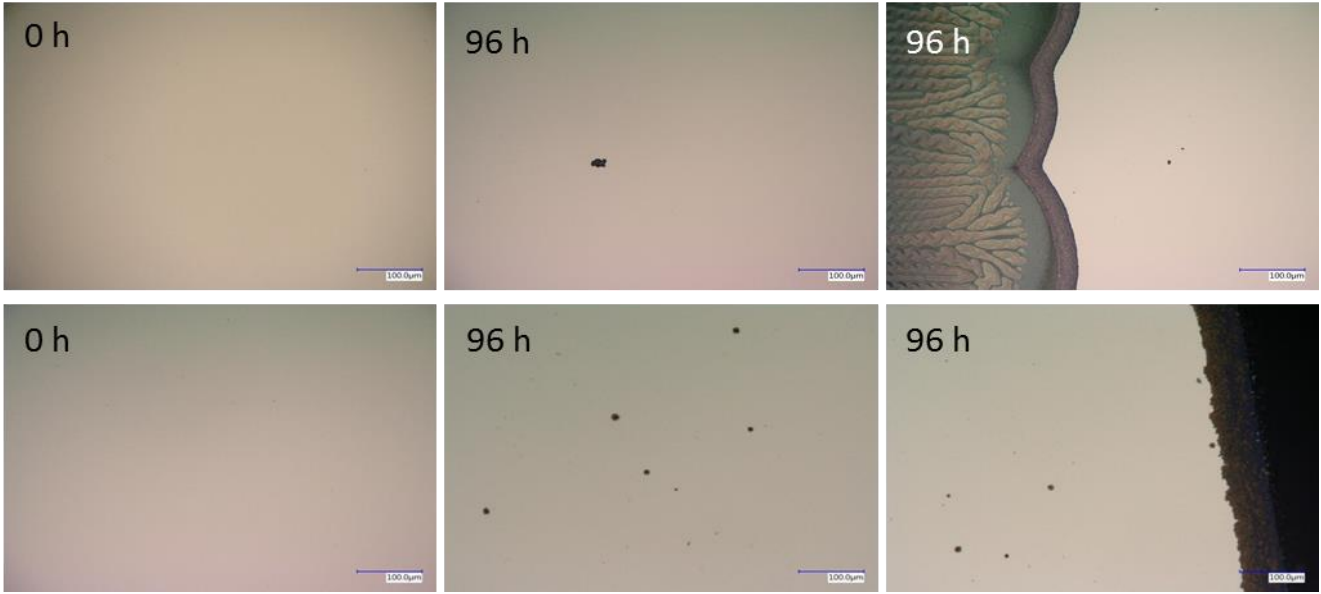

**Figure 5**

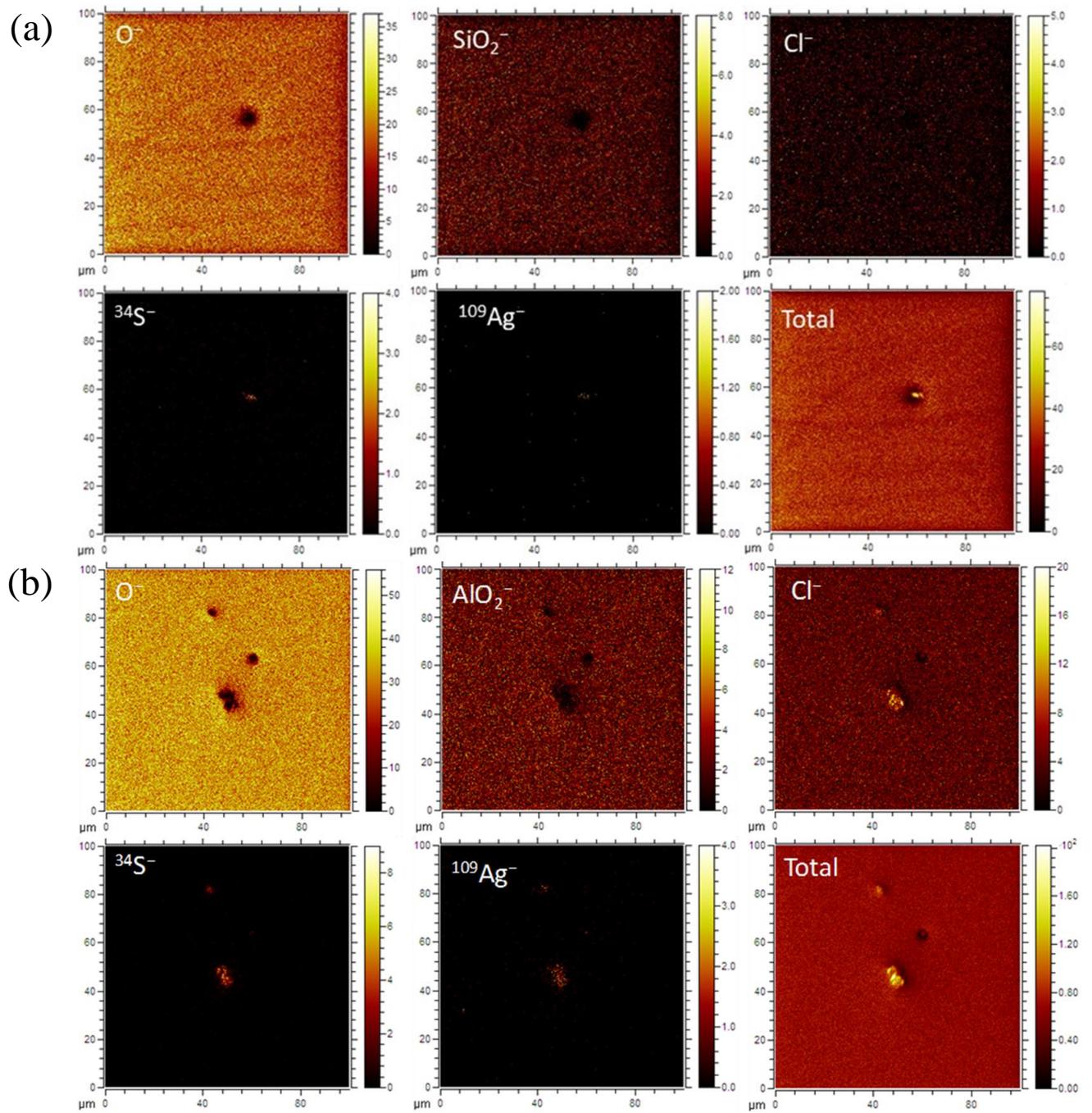

# Figure 6

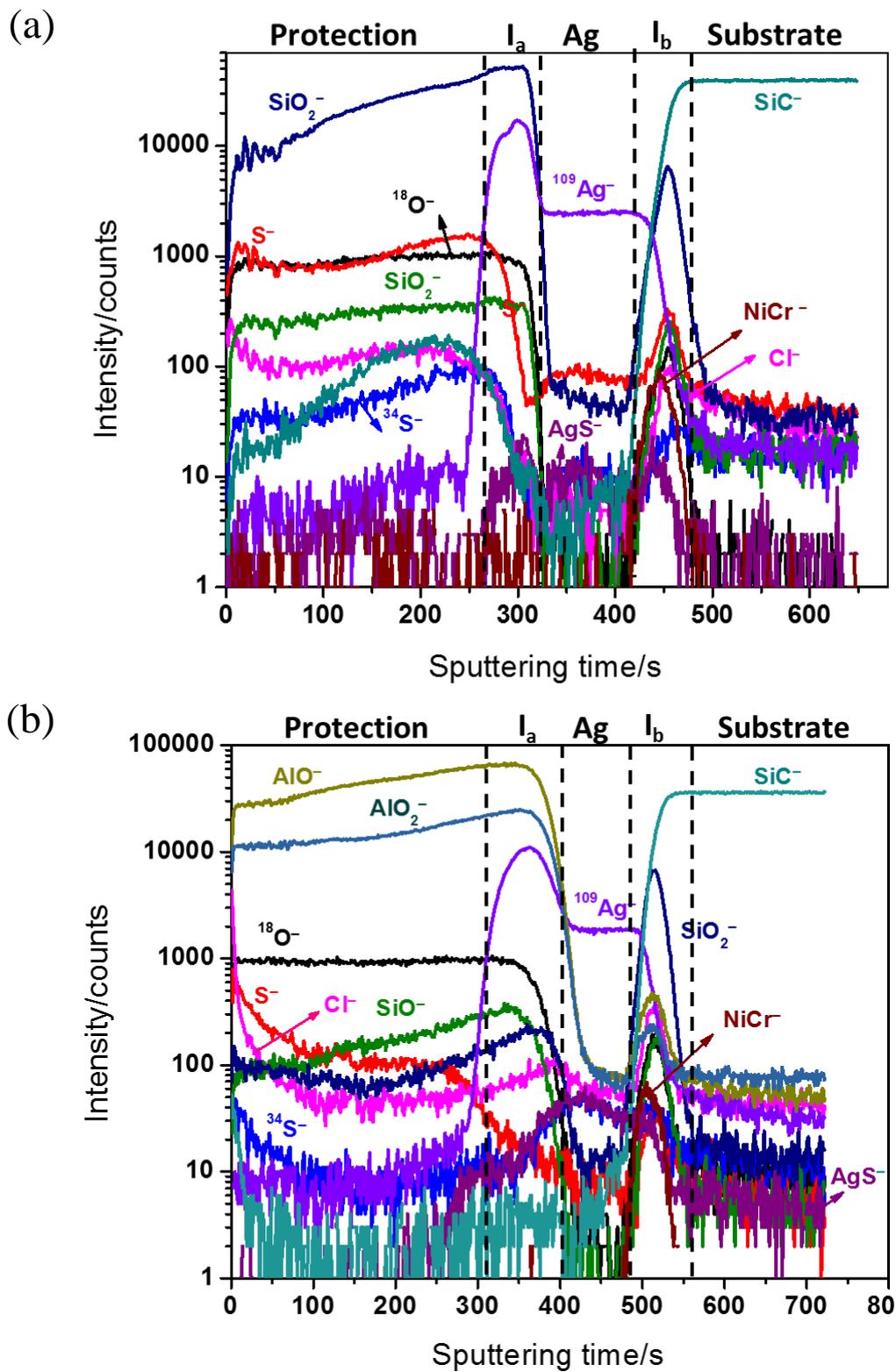

**Figure 7**

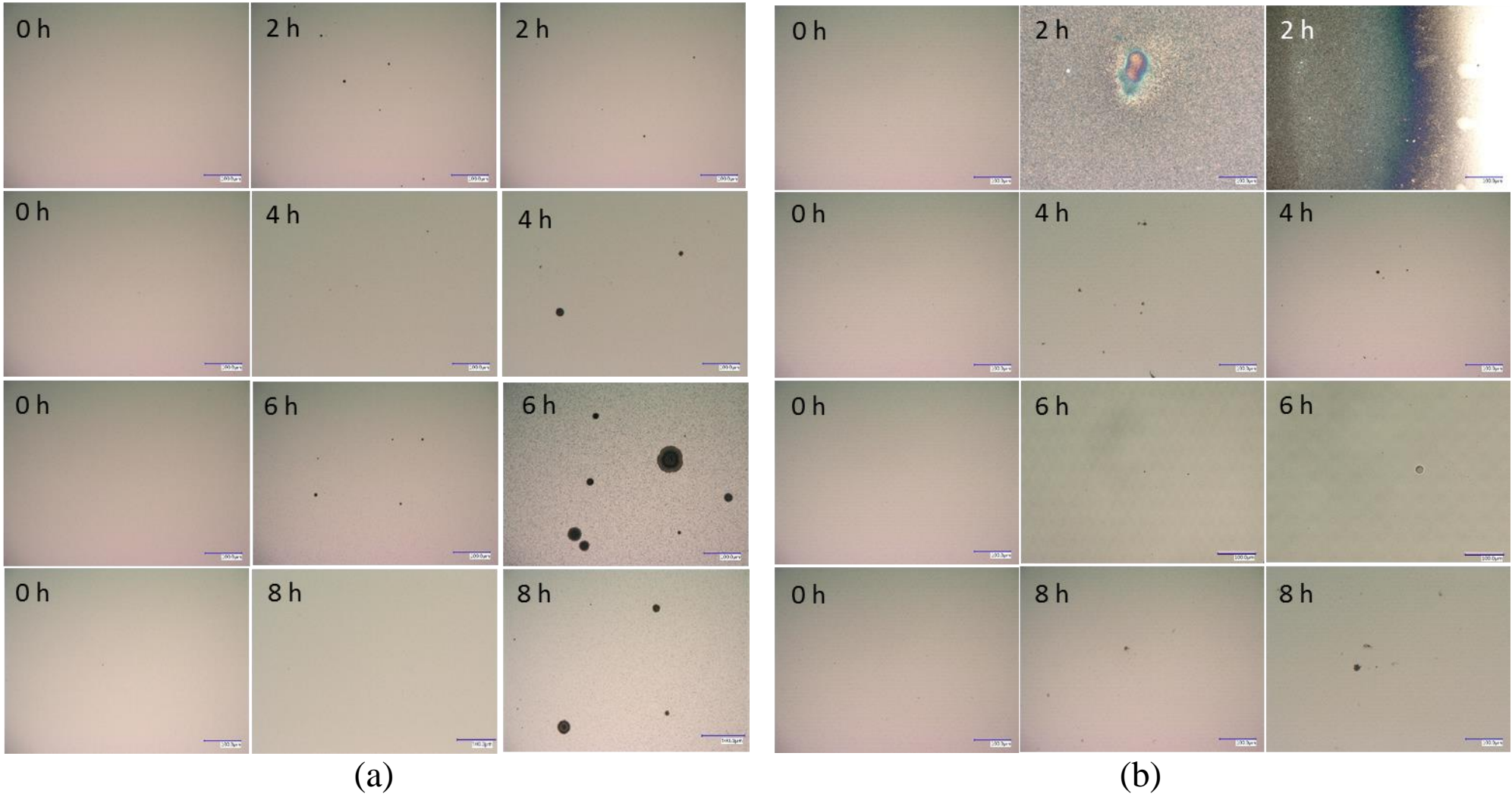

(a)                                        (b)

**Figure 8**

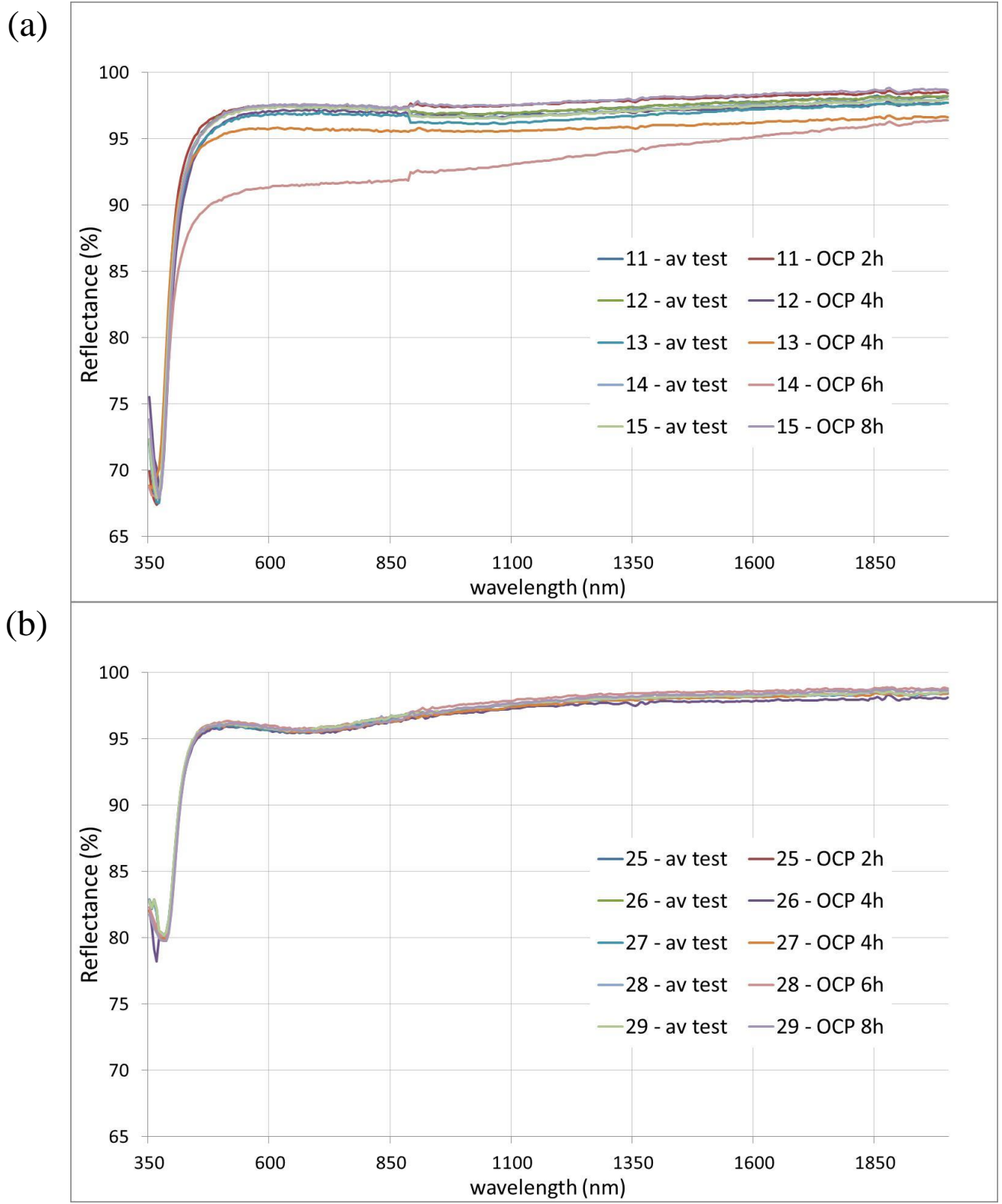